%% file: main.tex
\documentclass[sigconf]{acmart}

\copyrightyear{2024}
\acmYear{2024}
\setcopyright{acmlicensed}\acmConference[WSDM '24]{Proceedings of the 17th ACM International Conference on Web Search and Data Mining}{March 4--8, 2024}{Merida, Mexico}
\acmBooktitle{Proceedings of the 17th ACM International Conference on Web Search and Data Mining (WSDM '24), March 4--8, 2024, Merida, Mexico}
\acmPrice{15.00}
\acmDOI{10.1145/3616855.3635791}
\acmISBN{979-8-4007-0371-3/24/03}

\usepackage{booktabs}
\usepackage{graphicx}
\usepackage{multirow}
\usepackage{caption}
\usepackage{subcaption}
\input{math_commands.tex}

\begin{document}

%\title{Embedding-based Retrieval Models with \\Fast Semi-parametric Inference in Product Search}
%\title{\ERMada: Fast Adapter for Parameter-free Fine-tuning of Embedding-based Retrieval Models}
\title{PEFA: Parameter-Free Adapters for Large-scale Embedding-based Retrieval Models}

%\title{Lightweight Adapter for Parameter-free Fine-tuning of Bi-encoders in Large-scale Retrieval}

% \affiliation{%
%   \institution{Anonymous Institute}
%   \city{Anonymous}
%   \state{Anonymous}
%   \country{Anonymous}
% }
\settopmatter{authorsperrow=4}

\author{Wei-Cheng Chang}
\email{weicheng.cmu@gmail.com}
\affiliation{%
  \institution{Amazon}
  \city{Seattle}
  \state{Washington}
  \country{USA}
}

\author{Jyun-Yu Jiang}
\email{jyunyu.jiang@gmail.com}
\affiliation{%
  \institution{Amazon}
  \city{Palo Alto}
  \state{CA}
  \country{USA}
}

\author{Jiong Zhang}
\email{zhangjiong724@gmail.com}
\affiliation{%
  \institution{Amazon}
  \city{Palo Alto}
  \state{CA}
  \country{USA}
}

\author{Mutasem Al-Darabsah}
\email{mutasema@amazon.com}
\affiliation{%
  \institution{Amazon}
  \city{Palo Alto}
  \state{CA}
  \country{USA}
}

\author{Choon Hui Teo}
\email{choonhui@amazon.com}
\affiliation{%
  \institution{Amazon}
  \city{Palo Alto}
  \state{CA}
  \country{USA}
}

\author{Cho-Jui Hsieh}
\email{chohsieh@cs.ucla.edu}
\affiliation{%
  \institution{UCLA}
  \city{Los Angeles}
  \state{CA}
  \country{USA}
}

\author{Hsiang-Fu Yu}
\email{rofu.yu@gmail.com}
\affiliation{%
  \institution{Amazon}
  \city{Palo Alto}
  \state{CA}
  \country{USA}
}

\author{S. V. N. Vishwanathan}
\email{vishy@amazon.com}
\affiliation{%
  \institution{Amazon}
  \city{Palo Alto}
  \state{CA}
  \country{USA}
}

\renewcommand{\shortauthors}{Chang et al.}

\begin{abstract}
\input{contents/s0-abstract}
\end{abstract}

\ccsdesc[500]{Computing methodologies~Supervised learning}
\ccsdesc[500]{Computing methodologies~Instance-based learning}
\keywords{embedding-based retrieval, bi-encoders, parameter-free adapters, k-nearest neighbor models}
% %% A "teaser" image appears between the author and affiliation
% %% information and the body of the document, and typically spans the


\maketitle

\input{contents/s1-introduction}

\input{contents/s2-preliminary}

\input{contents/s3-method}

\input{contents/s4-exp-retrieval}
\input{contents/s5-exp-matching}
\input{contents/s6-relatedwork}
\input{contents/s7-conclusion}

\section*{Ethical Considerations}
We discuss ethical implications of our \PEFA framework in two perspectives: interpretability and privacy.

\paragraph{\bf Interpretability} 
Given a query, embedding-based retrieval models (\ERMs) retrieve the match set based on similarity search between the query embedding and the corpus of passage embeddings.
However, the interpretability and explainability of \ERMs is quite limited because we do not know which training examples contribute to or lead to the decisions of the retrieved match-set.
Our proposed framework \PEFA combines \ERMs with a non-parametric \kNN component, which enhances the interpretability of \ERMs.
The \kNN component computes similarity scores between the test query and the set of training queries, hence we know which training examples contribute the most to the retrieved match-set. 

\paragraph{\bf Privacy}
For e-commerce product search, it is crucial to protect customers privacy.
Thus, we need to insure the underlying models do not explicitly memorize customers purchase history.
When applying \PEFA to the product search datasets, we carefully anonymized the search log, hence we never know which customer issues a specific query.
Furthermore, we consider yearly-aggregated data of query-product pairs as the training signals in our \kNN component.
In other words, each query in our training set can not be traced back to its original query session.
%This can protect the privacy of each customer purchase history.

% \begin{acks}
% \end{acks}

\bibliographystyle{ACM-Reference-Format}
\balance
\bibliography{midas}

%\clearpage
%\appendix
%\input{appendix/appendix.tex}

\end{document}

%% file: math_commands.tex
\usepackage{xspace}
\usepackage{amsmath,amsfonts,bm}

\newcommand{\ERM}{ERM\xspace}

\newcommand{\ERMs}{ERMs\xspace}
\newcommand{\kNN}{kNN\xspace}
\newcommand{\ANN}{ANN\xspace}
\newcommand{\PEFA}{PEFA\xspace}
\newcommand{\PEFAxs}{PEFA-XS\xspace}
\newcommand{\PEFAxl}{PEFA-XL\xspace}

\newcommand{\SBERT}{$\text{Sent-BERT}_{\text{distill}}$\xspace}
\newcommand{\DPR}{$\text{DPR}_{\text{base}}$\xspace}
\newcommand{\MPNet}{$\text{MPNet}_{\text{base}}$\xspace}
\newcommand{\STfive}{$\text{Sentence-T5}_{\text{base}}$\xspace}
\newcommand{\GTR}{$\text{GTR}_{\text{base}}$\xspace}
\newcommand{\Efive}{$\text{E5}_{\text{base}}$\xspace}
\newcommand{\ProdERM}{FT-ERM\xspace}

\newcommand{\BM}{BM-25\xspace}
\newcommand{\NCI}{NCI\xspace}
\newcommand{\DSI}{DSI\xspace}
\newcommand{\SEAL}{SEAL\xspace}

\newcommand{\NQ}{NQ-320K\xspace}
\newcommand{\TriviaQA}{Trivia-QA\xspace}
\newcommand{\DATAca}{ProdSearch-30M\xspace}
\newcommand{\DATAmx}{ProdSearch-15M\xspace}
\newcommand{\DATAau}{ProdSearch-5M\xspace}
\def\eqref#1{equation~\ref{#1}}
\def\1{\bm{1}}

\def\rmD{{\mathbf{D}}}

\def\rmP{{\mathbf{P}}}
\def\rmQ{{\mathbf{Q}}}

\def\rmY{{\mathbf{Y}}}

\def\vp{{\bm{p}}}
\def\vq{{\bm{q}}}

\def\vx{{\bm{x}}}
\def\vy{{\bm{y}}}

\DeclareMathAlphabet{\mathsfit}{\encodingdefault}{\sfdefault}{m}{sl}
\SetMathAlphabet{\mathsfit}{bold}{\encodingdefault}{\sfdefault}{bx}{n}

\def\gD{{\mathcal{D}}}

\def\gI{{\mathcal{I}}}

\def\gO{{\mathcal{O}}}
\def\gP{{\mathcal{P}}}
\def\gQ{{\mathcal{Q}}}

\def\gS{{\mathcal{S}}}

\def\gX{{\mathcal{X}}}

\def\sR{{\mathbb{R}}}

%% file: contents/s0-abstract.tex
%Given a query, semantic matching models in product search retrieve a match set of relevant products in real time from an enormous catalog.
%Recently, embedding-based retrieval models (\ERMs) have emerged as a promising framework for solving semantic matching of product search.
Embedding-based Retrieval Models (\ERMs) have emerged as a promising framework for large-scale text retrieval problems due to powerful large language models.
Nevertheless, fine-tuning \ERMs to reach state-of-the-art results can be expensive due to the extreme scale of data as well as the complexity of multi-stages pipelines (e.g., pre-training, fine-tuning, distillation).
In this work, we propose the \PEFA framework, namely \textbf{P}aram\textbf{E}ter-\textbf{F}ree \textbf{A}dapters,
for fast tuning of \ERMs without any backward pass in the optimization.
At index building stage, \PEFA equips the \ERM with a non-parametric $k$-nearest neighbor (\kNN) component.
At inference stage, \PEFA performs a convex combination of two scoring functions, one from the \ERM and the other from the \kNN.
Based on the neighborhood definition, \PEFA framework induces two realizations, namely \PEFAxl (i.e., extra large) using double \ANN indices and \PEFAxs (i.e., extra small) using a single \ANN index.
Empirically, \PEFA achieves significant improvement on two retrieval applications.
%, including document retrieval and e-commerce product search.
For document retrieval, regarding Recall@100 metric, \PEFA improves not only pre-trained \ERMs on \TriviaQA by an average of 13.2\%, but also fine-tuned \ERMs on \NQ by an average of 5.5\%, respectively.
For product search, \PEFA improves the Recall@100 of the fine-tuned \ERMs by an average of 5.3\% and 14.5\%, for \PEFAxs and \PEFAxl, respectively.
Our code is available at~\url{https://github.com/amzn/pecos/tree/mainline/examples/pefa-wsdm24}.

%% file: contents/s1-introduction.tex
\section{Introduction}
%We consider the problem of large-scale text retrieval,
%which has huge impact to various real-world applications,
%such as search engines and e-commerce product search.
Given a user's query, large-scale text retrieval aims to recall a match set of semantically relevant documents in real-time from an enormous corpus, whose size can be 100 millions or more.
Embedding-based retrieval models (\ERMs)~\citep{karpukhin2020dense,chang2020pretraining,xiong2021approximate}, namely bi-encoders~\citep{nigam2019semantic,huang2020embedding}, have emerged as the prevalent paradigm for large-scale text retrieval, thanks to recent advances in large language models (LLMs).
At the learning stage, \ERMs fine-tune parametric Transformer encoders that map queries and documents into a semantic embedding space where relevant (query, document) pairs are close to each other and vice versa.
At the inference stage, retrieving relevant documents from the enormous output space can be formulated as the maximum inner product search (MIPS) problem~\citep{yu2017greedy}.
With proper indexing data structures, MIPS problem can be efficiently solved by approximate nearest neighbor (\ANN) search libraries (e.g., Faiss~\citep{johnson2019billion}, ScaNN~\citep{guo2020accelerating}, HNSWLIB~\citep{malkov2018efficient}) in time sub-linear to the size of corpus.

Adapting \ERMs to downstream retrieval tasks usually follows the full-parameter fine-tuning paradigm, which requires gradient computations and updates parameters of Transformer encoders.
Such full-parameter fine-tuning approach faces challenges in the industrial setup, where learning signals are enormous.
In modern e-commerce stores, for example, the number of relevant (query, product) pairs can be billions or more.
Full-parameter fine-tuning \ERMs on such scale may take thousands of GPU hours due to complicated multi-stage pipeline: 
pre-training~\citep{chang2020pretraining,gao2021condenser,gao2022unsupervised}, 1st stage fine-tuning with random negatives and BM25 candidates~\citep{karpukhin2020dense}, 2nd stage fine-tuning with hard-mined negatives~\citep{xiong2021approximate,zhan2021optimizing}, and 3rd stage fine-tuning with distilled knowledge from expensive cross-attention models~\citep{ren2021rocketqav2,zhang2022adversarial}.
Furthermore, these fine-tuning approaches require access to models' gradient information, which is not accessible for many black box LLMs such as GPT-3~\citep{brown2020language} and beyond.

In this work, we propose the \PEFA framework (i.e., \textbf{P}aram\textbf{E}ter-\textbf{F}ree \textbf{A}dapters) for fast tuning of black-box \ERMs, which doesn't require any gradient information of the model.
The scoring function of \PEFA is a convex combination
between the \ERM and the new non-parametric $k$-nearest neighbor (\kNN) model.
The learning of \kNN model reduces to constructing \ANN index
that stores key-value pairs of query embeddings and learning signals.
Given a query at inference time,
the \kNN model seeks close-by training queries in the neighborhood,
and aggregates the associated relevant documents as its scoring function.
Depending on the definition of neighborhood, we introduce two \kNN models under our \PEFA framework:
\begin{itemize}
    \item \PEFAxs: the neighborhood is defined by the relevant query-document pairs, which is independent to the test-time query.
    \item \PEFAxl: the neighborhood is an intersection of the one in \PEFAxs and \kNN queries in the training set, which is dependent to the test-time query.
\end{itemize}
In Summary, we highlight four key contributions below.
\begin{itemize}
    \item We propose \PEFA, a novel \textit{Parameter-free} adapters framework for fast tuning \ERMs to downstream retrieval tasks.
    \item \PEFA requires no gradient information of \ERMs, hence applicable to black-box \ERMs.
    \item \PEFA is not only applicable to a wide-range of pre-trained \ERM, but also effective to fine-tuned \ERMs.
    \item We demonstrate the effectiveness and scalability of \PEFA on two retrieval applications,
    including document retrieval tasks and industrial-scale product search tasks.
\end{itemize}
For document retrieval,
\PEFA not only improves the recall@100 of pre-trained \ERMs on \TriviaQA by an average of 13.2\%,
but also lifts the recall@100 of fine-tuned \ERMs on \NQ by an average of 5.5\%.
For \NQ dataset, applying \PEFA to the fine-tuned \GTR~\citep{ni2022large} reaches new state-of-the-art (SoTA) results,
where the Recall@10 of $88.71\%$ outperforms the Recall@10 of $85.20\%$ in the previous SoTA Seq2Seq-based \NCI~\citep{wang2022neural},
under similar model size for fair comparison.
For product search consisting of billion-scale of data,
\PEFA improves the Recall@100 of the fine-tuned \ERMs
by an average of 5.3\% and 14.5\%, for \PEFAxs and \PEFAxl, respectively.

%% file: contents/s2-preliminary.tex
\section{Preliminary}
\label{sec:preliminary}

\subsection{Dense Text Retrieval}
\label{sec:erm}
Dense text retrieval typically adopts the Embedding-based Retrieval Model (\ERM) architecture,
also known as bi-encoders~\citep{karpukhin2020dense,xiong2021approximate,chang2020pretraining}.
For simplicity, we use the term passage/document interchangeably in the rest of paper.
Given a query $q \in \gX$ and a passage $p \in \gX$,
the relevance scoring function $f_{\text{\ERM}}(q, p)$ of the \ERM
is measured by
%$f_{\text{\ERM}}(q, p; \theta) = \bigl< E(q; \theta), E(p; \theta) \bigr>$
\begin{equation}
    f_{\text{\ERM}}(q, p; \theta)
    = \bigl< E(q; \theta), E(p; \theta) \bigr>,
    \label{eq:erm}
\end{equation}
where $E(\cdot; \theta): \gX \rightarrow \sR^d$ is the encoder parameterized with $\theta$ that maps an input text to a  $d$-dimensional vector space and $\langle \cdot, \cdot \rangle: \sR^d \times \sR^d \rightarrow \sR$ is the similarity function, including inner product and cosine similarity.
Without loss of generality, we use inner product as the scoring function for the rest of paper.
%Next, we discuss the learning (i.e., how to estimate the encoder parameters $\theta$) and the inference (i.e., how to do real-time prediction at scale) of \ERMs.
%Next, we discuss the learning and the inference of \ERMs.

\paragraph{\bf Learning}
Suppose the training data is presented as a set of relevant query-passage pairs $\gD=\{(q_i, p_i)\}_{i=1}^{|\gD|}$.
The encoder parameters $\theta$ are often learned by maximizing the log-likelihood loss function ~\citep{karpukhin2020dense,chang2020pretraining}
%\begin{equation*}
    $\max_{\theta}  \sum_{(q,p) \in \gD} \log p_{\theta}(p|q)$,
%\end{equation*}
where the conditional probability is defined by the Softmax function
\begin{equation*}
     p_{\theta}(p|q)
    = \frac{\exp{ \big( f_{\text{\ERM}}(q,p;\theta) \big) }}
    {\sum_{p'\in\gD} \exp{ \big( f_{\text{\ERM}}(q,p';\theta) \big) }}.
\end{equation*}
In practice, various negative sampling techniques~\citep{karpukhin2020dense,xiong2021approximate,lin2021batch,formal2022distillation} have been developed to approximate the expensive partition function of the conditional Softmax.
We direct interested readers to the comprehensive study~\cite{guo2022semantic} for more details of learning \ERMs.

\paragraph{\bf Inference}
Given a query embedding $\vq \in \sR^d$ and
a corpus of $n$ passage embeddings $\gP=\{ \vp_j\}_{j=1}^n$ where $\vp_j \in \sR^d, j=1,\ldots, n$,
\ERMs retrieve $k$ most relevant passages from $\gP$ in real time, which is a Maximum Inner Product Search (MIPS) problem.
% \begin{equation}
%     \argmax_{j=1,\ldots,n} \bigl< \vq, \vp_{j} \bigr>.
% \end{equation}
Exact inference of MIPS problem requires $O(n)$ time, which is prohibited for large-scale retrieval applications.
Thus, practitioners leverage Approximate Nearest Neighbor search (\ANN)
to approximately solve it in time \textit{sub-linear} (e.g., $\gO\big(\log(n)\big)$) to the size of corpus $n$.

To achieve sub-linear time complexity of \ANN search,
\ANN methods require an additional \textit{index building stage} to preprocess the corpus $\gP$ into specific data structures,
such as hierarchical graphs (e.g., HNSW~\citep{malkov2018efficient}, VAMANA~\cite{jayaram2019diskann}, etc)
and product quantization (e.g., FAISS~\citep{johnson2019billion}, ScaNN~\citep{guo2020accelerating}, etc).
Compared to the cost of full-parameter fine-tuning \ERMs on GPU machines,
the cost of building ANN index is often negligible 
as the latter takes place on the lower-cost single CPU machine, with much faster computational time.

\subsection{Problem Statement}

\begin{figure*}[!htb]
    \centering
    \begin{minipage}{0.49\textwidth}
        \centering
        \includegraphics[width=0.87\linewidth]{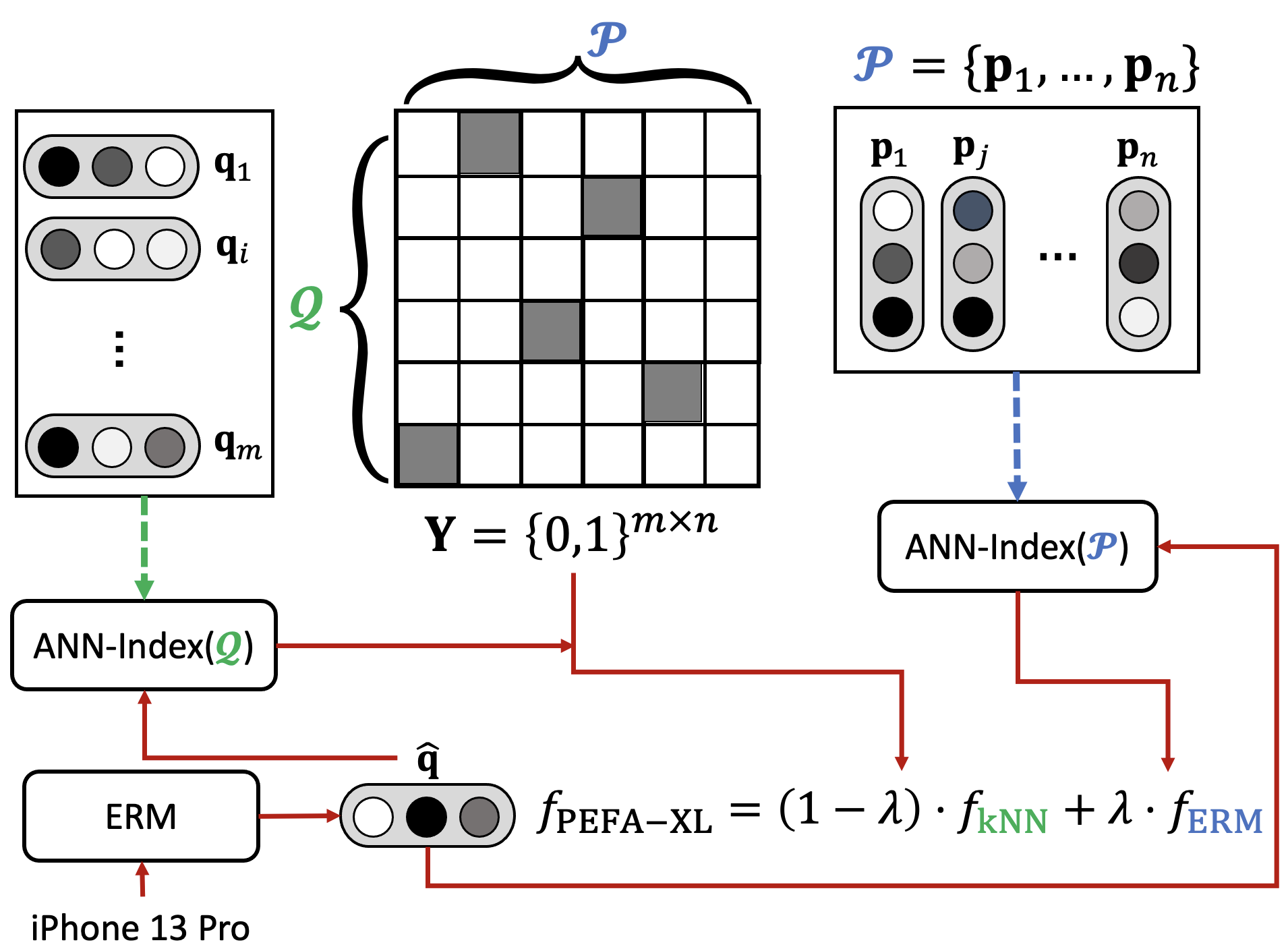}
        \caption{
            Illustration of the proposed \PEFAxl method. 
            The green and blue dash line represents computation paths for building \ANN index in \textit{offline}.
            The red solid line represents computation paths for performing ANN search in \textit{online}. 
            \PEFAxl requires \textit{two} \ANN indices:
            one on the passage space $\gP$ for fast inference of \ERM, while the other on the query space $\gQ$ for fast inference of \kNN model. 
        }
        \label{fig:pefa-xl}
    \end{minipage}%
    \hfill
    \begin{minipage}{0.49\textwidth}
        \centering
        \includegraphics[width=0.92\linewidth]{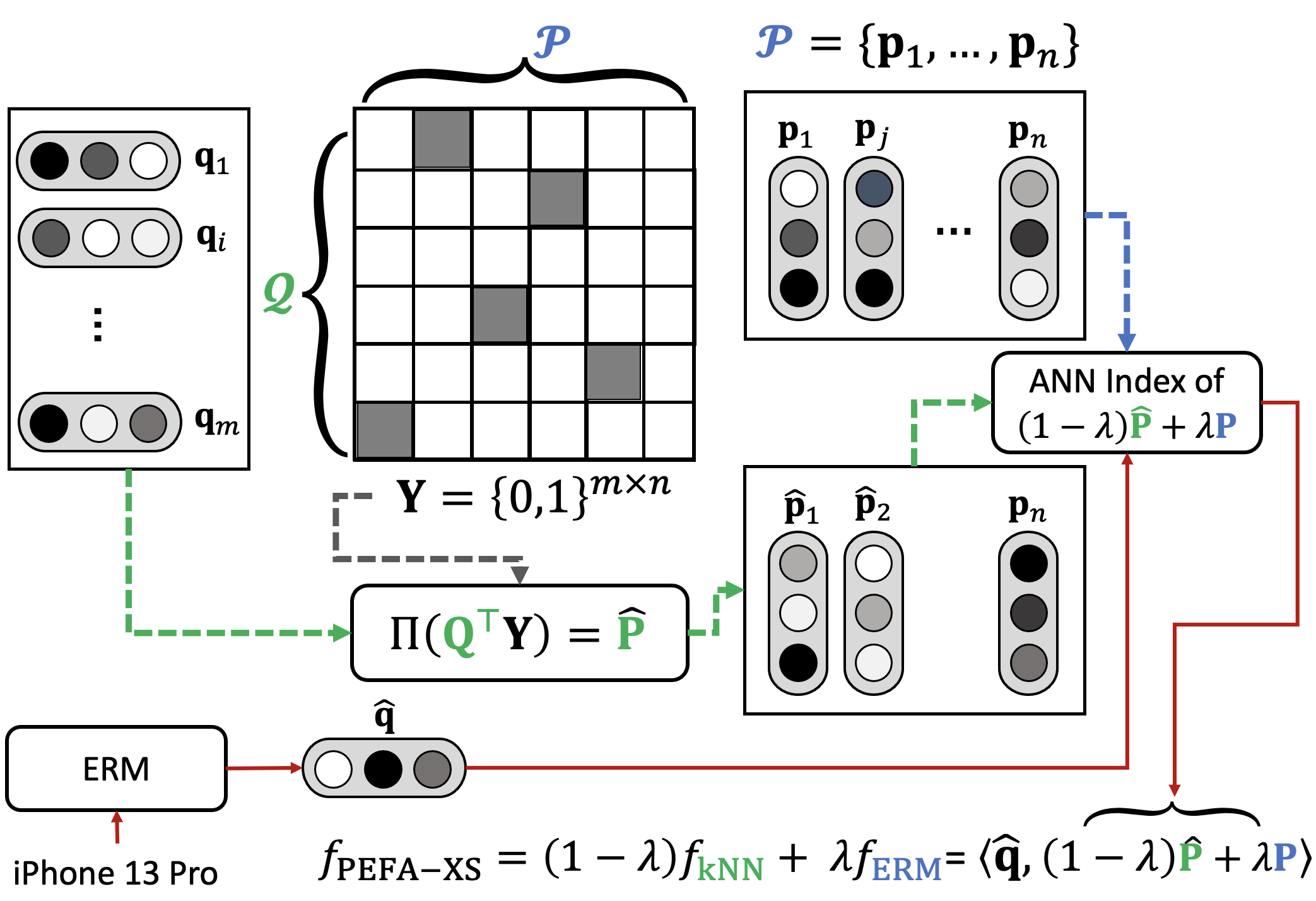}
        \caption{
            Illustration of the proposed \PEFAxs method. 
            The green and blue dash line represents computation paths for building ANN index in \textit{offline}.
            The red solid line represents computation paths for performing ANN search in \textit{online}.
            \PEFAxs requires only \textit{one} \ANN index.
            As the neighborhood of \kNN becomes independent to $\hat{\vq}$, interpolation of two score functions can be pre-computed in the embedding space,
            when building the ANN index \textit{offline}.
        }
        \label{fig:pefa-xs}
    \end{minipage}
\end{figure*}

\paragraph{\bf Notations.}
%We first setup notations for the rest of this paper.
$\rmY \in \{0,1\}^{m \times n}$ is the query-to-passage relevant matrix, namely the supervised training data.
%For simplicity, we assume binary relevance of query-passage pairs in this paper.
The row indices of $\rmY$ refer to the set of queries $\gQ$,
and the column indices of $\rmY$ refer to the set of passages $\gP$, respectively.
Note that $\gP$ is also the corpus space used for ANN inference of \ERMs.
The bold matrices $\rmP \in \sR^{n \times d}$ and $\rmQ \in \sR^{m \times d}$
denote the query and passage embeddings of $\gP$ and $\gQ$ obtained from the \ERM, respectively.
We denote $\rmY_{i,:} \in \{0,1\}^{n}$ as the relevant vector of $i$th query $\vq_i$,
representing which set of passages in $\gP$ are relevant to this query $\vq_i$.
Similarly, $\rmY_{:,j} \in \{0,1\}^{m}$ is the relevant vector of $j$th passage $\vq_j$,
representing which set of queries in $\gQ$ are relevant to this passage $\vp_j$.
Finally, $\hat{\vq} \in \sR^d$ is the query embeddings at inference time and 
$\text{NN}(\hat{\vq},\rmP;k)$ is the set of $k$ nearest indices in the indexed database $\rmP$ given $\hat{\vq}$.
%The notations commonly-used in the rest of paper is also summarized in Table~\ref{tab:notation} of Appendix~\ref{apx:notations}.

\paragraph{\bf Problem Setup.}
In this work, we propose \PEFA, \textit{parameter-free} adapters for \ERMs via equipping it with a non-parametric \kNN component.
The non-parametric \kNN model is learning-free which avoid any optimization step to fine-tune parameters of \ERMs.
The major computation of \PEFA becomes building \ANN index storing key-value pairs for serving efficiently at the inference stage.
Thus, our \PEFA is also applicable to both pre-trained and fine-tuned \ERMs, even ones initialized from black-box LLMs.
%From the development cycles (learning and inference) of \ERMs described in Section~\ref{sec:erm},
%the scope of this paper falls into the \textit{inference stage} (i.e., ANN index building and serving).
Note that \PEFA is orthogonal and complement to most existing literature that aims to obtain better pre-trained or fine-tuned \ERMs at the learning stage,
including recent studies of the parameter-efficient fine-tuning of \ERMs~\citep{jung2022semi,ma2022scattered,pal2023parameter}.
Finally, for the ease of discussion, we assume embeddings obtained from \ERMs are unit-norm (i.e., $\ell_2$ normalized),
hence the inner product is equivalent to the cosine similarity.
The techniques proposed in this paper can be easily extended to non-unit norm cases by replacing the distance metric used in \kNN.

%% file: contents/s3-method.tex
\section{Proposed Framework}
\label{sec:proposed}

In this section, we propose \PEFA,
a \textit{Parameter-free} Adapters framework for fast tuning of \ERMs.
Given a query embedding $\hat{\vq}$ at inference time,
our \PEFA framework defines the relevant scoring function of a query-passage pair $(\hat{\vq}, \vp_j)$
as the convex combination between scoring functions of the black-box \ERM and a non-parametric \kNN model:
\begin{equation}
    f_{\text{\PEFA}}(\hat{\vq}, \vp_j)
    = \lambda \cdot f_{\text{\ERM}}(\hat{\vq}, \vp_j)
    + (1-\lambda) \cdot f_{\text{\kNN}}(\hat{\vq}, \vp_j),
    \label{eq:pefa-general}
\end{equation}
where $\lambda \in [0, 1]$ is the interpolation hyper-parameter to balance the importance between the \ERM and the \kNN model. 
Note that the proposed \PEFA framework is learning-free.
In other words,
the underlying parameters $\theta$ of \ERM remains unchanged,
and Equation~\ref{eq:pefa-general} is only applied at the inference time.

%\peter{Some context about \kNN models here.}
Next, we present scoring functions of \kNN models in a generic form as follows.
\begin{equation}
f_{\text{\kNN}}(\hat{\vq}, \vp_j)
    = \bigl< \hat{\vq}, \rmQ^\top \rmD(\hat{\vq}, \rmQ) \rmY_{:,j} \bigr> 
    \label{eq:knn-mat}
\end{equation}
where $\rmD(\hat{\vq}, \rmQ) \in \sR^{n \times n}$ is a normalized diagonal matrix,
acting like a gating mechanism that controls which set of training queries the current test query $\hat{\vq}$ should pay attention to. 

Plugging Equation~\ref{eq:knn-mat} back to Equation~\ref{eq:pefa-general},
we derive the scoring function of \PEFA explicitly
\begin{equation}
    f_{\text{\PEFA}}(\hat{\vq}, \vp_j)
    = \lambda \bigl< \hat{\vq}, \vp_j \bigr>
    + (1-\lambda) \bigl< \hat{\vq}, \rmQ^\top \rmD(\hat{\vq}, \rmQ) \rmY_{:,j} \bigr>.
    \label{eq:pefa-explicit}
\end{equation}

Based on the design of diagonal matrix $\rmD(\hat{\vq}, \rmQ)$,
we present two realizations of \kNN models under the \PEFA framework,
namely \PEFAxl (Section~\ref{sec:pefa-xl}) and \PEFAxs (Section~\ref{sec:pefa-xs}).
We then discuss their intuitions, time and space complexity, and connections to the related literature.
For the rest of the paper,
We use HNSW~\citep{malkov2018efficient} as the underlying \ANN methods in our \PEFA framework for complexity analysis and experiment results.

\subsection{\PEFAxl}
\label{sec:pefa-xl}

A standard realization of the \kNN model is that the test query $\hat{\vq}$ only pays attention to top-$k$ most similar training queries in $\gQ$.
Specifically, $\rmD_{i,i}=1$ if $i \in \text{NN}(\hat{\vq}, \rmQ; k)$; otherwise $\rmD_{i,i}=0$.
We can then derive the \kNN model of Equation~\ref{eq:knn-mat} as following:
\begin{equation}
f_{\text{\kNN}}(\hat{\vq}, \vp_j)
    = \bigl< \hat{\vq}, \sum_{i=1}^n (\rmD_{i,i} \rmY_{i,j}) \cdot \vq_i \bigr>
    = \sum_{i \in \text{NN}(\hat{\vq},\rmQ;k)} 
    \bigl< \hat{\vq}, \vq_i \bigr> \cdot \rmY_{i,j}.
    \label{eq:knn-xl}
\end{equation}
By plugging the query-aware \kNN model of Equation~\ref{eq:knn-xl} into Equation~\ref{eq:pefa-general},
we present the scoring function of \PEFAxl explicitly
\begin{equation}
    f_{\text{\PEFAxl}}(\hat{\vq}, \vp_j)
    = \lambda \bigl< \hat{\vq}, \vp_j \bigr>
    + (1-\lambda) \sum_{i \in \text{NN}(\hat{\vq},\rmQ;k)} 
    \bigl< \hat{\vq}, \vq_i \bigr> \cdot \rmY_{i,j}.
    \label{eq:pefa-xl}    
\end{equation}

\paragraph{\bf Implementation.}
An illustration of \PEFAxl method is presented in Figure~\ref{fig:pefa-xl}.
Intuitively, the \kNN model of \PEFAxl produces its match set
by aggregating relevant passages $\rmY_{i,j}$ of training queries $\vq_i$ that are in the neighborhood of the test query $\hat{\vq}$.
Note that the scoring function of $f_{\text{\ERM}}$ in Equation~\ref{eq:pefa-xl}
is bounded between $[-1, 1]$ as the inner product of two unit-form embeddings are bounded by the range of cosine similarity.
On the other hand, the scoring function of $f_{\text{\kNN}}$ in Equation~\ref{eq:pefa-xl}
needs an additional normalization so that its score is calibrated to $f_{\text{\ERM}}$.
In practice, we consider normalizing $f_{\text{\kNN}}$ by $k$, namely $D_{i,i} = 1 / k$ if $i \in NN(\hat{\vq},\rmQ; k)$,
so that $f_{\text{\kNN}}$ is still upper bounded by $1$.

\begin{table*}[!ht]
    \centering
    \begin{tabular}{c|c|c|c|c}
        \toprule
        Methods & Scoring functions & HNSW index building time & HNSW index size & HNSW inference time \\
        \midrule
        \ERM    & Equation~\ref{eq:erm} & $\gO\big( n\log(n) \big)$ & $\gO\big( nd + |E_\gP| \big)$ & $\gO\big( \log(n) \big)$ \\
        \PEFAxl & Equation~\ref{eq:pefa-xl} & $\gO(n\log(n) + m\log(m))$ & $\gO\big( nd+|E_\gP|+md+|E_\gQ|+\text{nnz}(\rmY) \big)$ & $\gO\big( \log(n) + \log(m) \big)$ \\
        \PEFAxs & Equation~\ref{eq:pefa-xs} & $\gO\big( n\log(n) \big)$ & $\gO\big( nd + |E_\gP| \big)$ & $\gO\big( \log(n) \big)$ \\
        \bottomrule
    \end{tabular}
    \caption{Comparing time and space complexity of \ERM, \PEFAxs, and \PEFAxl at the inference stage.
    We use the competitive graph-based \ANN algorithm HNSW~\citep{malkov2018efficient} as the underlying method for \ANN search.
    The time and space complexity of HNSW is induced from ~\citep{wang2021comprehensive}.
    $E_\gP$ and $E_\gQ$ represents the edges of HNSW graph for $\gP$ and $\gQ$, respectively.}
    \label{tab:complexity}
    \vspace{-1em}
\end{table*}

\paragraph{\bf Complexity Analysis.}
At inference stage, retrieving the match set via \PEFAxl (Equation~\ref{eq:pefa-xl}) requires \ANN searches on \textbf{TWO} distinct output spaces.
Specifically, the \ERM requires \ANN search on the passage space $\gP$ of size $n$,
while the \kNN model requires \ANN search on the query space $\gQ$ of size $m$.
According to the comprehensive review~\citep{wang2021comprehensive},  the inference time complexity of HNSW on a data set $\gS$ is $\gO\big( \log(|\gS|) \big)$.
Thus, the inference time complexity of \PEFAxl becomes $\gO\big( \log(n) + \log(m) \big)$.

Next, we discuss the space complexity of \PEFAxl, which requires to store two HNSW indices as well as the query-to-passage relevant matrix $\rmY$.
The space complexity of an HNSW index on a data set $\gS$ is $\gO\big( |\gS|d + |E_\gS| \big)$ where the former comes from saving the database vectors and the latter comes from saving edges of the HNSW graph.
The space complexity of storing $\rmY$ is $\gO\big( \text{nnz}(\rmY) \big)$.
Thus, the space complexity of \PEFAxl is $\gO\big( nd + |E_\gP| + md + |E_\gQ| + \text{nnz}(\rmY) \big)$.

Finally, the time complexity of building HNSW indices for \PEFAxl is $\gO\big( n\log(n) + m\log(m) \big)$ because building a HNSW index for a set $\gS$ takes $\gO\big( |\gS|\log(|\gS|) \big)$~\citep{wang2021comprehensive}.
The building time, inference time, and space complexity of \PEFAxl is summarized in Table~\ref{tab:complexity}. 

\paragraph{\bf Connections to kNN-LM}
Using a non-parametric \kNN model to improve a parametric neural network has also been studied in the context of $k$ nearest neighbors language modeling (\kNN-LM)~\citep{khandelwal2020Generalization,yogatama2021adaptive,he2021efficient} and retrieval-augmented LM pre-training~\citep{guu2020retrieval,lewis2020retrieval,borgeaud2022improving}.
\kNN-LM interpolates the next-token predictive probability by the neural language model and the kNN model.
While sharing similar intuitions, \PEFAxl is different from \kNN-LM because \PEFAxl requires two ANN searches,
one on the passage space $\gP$ and the other on the query space $\gQ$.
In contrast, \kNN-LM only needs one ANN search on the context space, while inference on the vocab space is exact since the size of vocabulary is small.

\subsection{\PEFAxs}
\label{sec:pefa-xs}
In practice, \PEFAxl can be too expensive to deploy due to storing two \ANN indices, which double the model storage and inference latency. 
Thus, we seek an efficient alternative of \PEFAxl that only needs to maintain single \ANN index, hence the name \PEFAxs.

Recall that \PEFAxl demands another ANN search because of finding $k$ nearest queries in $\gQ$, namely $\text{NN}(\hat{\vq}, \gQ; k)$.
We can approximate $\text{NN}(\hat{\vq}, \gQ; k)$ by the set of relevant queries given a target passage $\vp_j \in \gP$.
We denote this alternative query set as $\gI(\vp_j, \rmY)=\{ i | \rmY_{ij} > 0, i=1,\ldots, n\}$, which is a function of $\vp_j$ that is independent to the test query $\hat{\vq}$.
In other words, the resulting diagonal matrix $\rmD_{i,i}=1$ if $i \in \gI(\vp_j, \rmY)$; other $\rmD_{i,i}=0$.
The approximate \kNN model of \PEFAxs becomes
\begin{equation}
f_{\text{\kNN}}(\hat{\vq}, \vp_j)
    = \bigl< \hat{\vq}, \sum_{i \in \gI(\vp_j, \rmY)} \rmY_{i,j} \cdot \vq_i \bigr>
    = \bigl< \hat{\vq}, \rmQ^\top \rmY_{:,j} \bigr>.
    \label{eq:knn-xs}
\end{equation}
By plugging the query-independent \kNN model of Equation~\ref{eq:knn-xs} back to Equation~\ref{eq:pefa-general}, we present the scoring function of \PEFAxs
\begin{equation*}
    f_{\text{\PEFAxs}}(\hat{\vq}, \vp_j)
    = \lambda \bigl< \hat{\vq}, \vp_j \bigr>
    + (1-\lambda) \bigl< \hat{\vq}, \rmQ^\top \rmY_{:,j} \bigr>.
\end{equation*}

\paragraph{\bf Implementation}
An illustration of \PEFAxs method is presented in Figure~\ref{fig:pefa-xs}.
Similar to the implementation design of \PEFAxl, we need to normalize the scoring function of $f_{\text{\kNN}}$ in Equation~\ref{eq:knn-xs} such that its score is upper bounded by $1$.
Thus, we introduce an $\ell_2$ normalization operator $\Pi(\vx)=\frac{\vx}{\|\vx\|}$ that projects an embedding back to the unit-sphere.
We can then rewrite the scoring function of \PEFAxs as
\begin{equation}
    f_{\text{\PEFAxs}}(\hat{\vq}, \vp_j)
    = \bigl< \hat{\vq}, \\
    \lambda\cdot\vp_j +(1-\lambda)\cdot \Pi(\rmQ^\top \rmY_{:,j}) \bigr>, \label{eq:pefa-xs}
\end{equation}
where normalization step $\Pi(\cdot)$ can be absorbed in the design of $\rmD$.
%, $\rmD_{i,i} = \frac{1.0}{\|\rmQ^\top \rmY_{:,j} \|}$.

\paragraph{\bf Complexity Analysis}
Note that the \kNN model of \PEFAxs is independent to the test query $\hat{\vq}$,
so the interpolation of two scoring functions can be pre-computed in the embedding space, as derived in Equation~\ref{eq:pefa-xs}.
This suggests that \PEFAxs only requires a single ANN index,
where a set of $n$ interpolated passage embeddings are used.
Therefore, the inference of \PEFAxs share the same time and space complexity as \ERM alone.
Specifically, the time complexity of constructing HNSW index and performing ANN search are $\gO\big( n \log(n) \big)$ and $\gO\big( \log(n) \big)$, respectively.
The space complexity of storing ANN index is $\gO(nd + |E_\gP|)$.
Finally, the time and space complexity of \PEFAxs is summarized in Table~\ref{tab:complexity}.

\paragraph{\bf Connections to XMC}
Given a passage, aggregating its relevant (as defined by customer behavior signals) query embeddings to be an alternative passage embeddings of itself has also been explored in the extreme multi-label classification (XMC) literature.
Specifically, XMC community terms such representation as \textbf{P}ostive \textbf{I}nstance \textbf{F}eature \textbf{A}ggregation, namely PIFA embeddings~\citep{jain2019slice,chang2021extreme,zhang2021fast,yu2022pecos}.
However, PIFA embeddings are often an aggregation of sparse tfidf features, and mostly used for unsupervised clustering to partition label space in the XMC literature~\citep{yu2022pecos}.
In contrast, \PEFAxs interpolates such alternative passage embeddings with the original passage embeddings, and conduct ANN search on the interpolated ASIN embedding space. 
As a side note, PIFA embeddings are also closely connected to the simple graph convolution layer in graph neural network~\citep{wu2019simplifying,chien2021node,lu2021graph}, where the input-to-label relevant matrix is viewed as a bipartite graph.

%% file: contents/s4-exp-retrieval.tex
%\newpage

\section{Experiments on Document Retrieval}
\label{sec:exp-retrieval}

In this section, we empirically verify the effectiveness of \PEFA on the document retrieval task. Experiment code is available at
~\url{https://github.com/amzn/pecos/tree/mainline/examples/pefa-wsdm24}.

\subsection{Datasets \& Evaluation Protocols}
\label{sec:exp-retrieval-setup}

\paragraph{\bf Datasets.}
We conducted experiments on two public benchmarks for document retrieval, namely
the Natural Questions~\citep{kwiatkowski2019natural} dataset and the \TriviaQA~\citep{joshi2017triviaqa} dataset.
\begin{itemize}
    \item Natural Questions~\citep{kwiatkowski2019natural}: a open-domain question answering dataset which consists of $320k$ query-document pairs, where the documents containing answers are gathered from Wikipedia and the queries are natural questions.
    The version we use is often referred to as \NQ~\citep{bevilacqua2022autoregressive,tay2022transformer,wang2022neural}.
    \item \TriviaQA~\citep{joshi2017triviaqa}: a reading comprehension dataset which includes $78k$ query-document pairs from the Wikipedia domain.
    We use the same version as in ~\citep{wang2022neural}.
\end{itemize}
The State-of-the-art (SoTA) method, namely \NCI~\citep{wang2022neural} consider generated query-document pairs as additional training signals in \NQ and \TriviaQA datasets.
Following the same setup of NCI~\citep{wang2022neural}, we also include those augmented query-document pairs when learning \PEFA on \NQ and \TriviaQA datasets.

\paragraph{\bf Evaluation Protocols}
We measure the performance with recall metrics,
which are widely-used in retrieval communities~\citep{karpukhin2020dense,chang2020pretraining, ni2022sentence,tay2022transformer,wang2022neural}.
Specifically, given a predicted score vector $\hat{\vy} \in \sR^{n}$ and a ground truth label vector $\vy \in \{0,1\}^n$, Recall@k is defined as $\text{Recall}@k = \frac{1}{|\vy|}\sum_{j \in \text{top}_k(\hat{\vy})} \vy_j$
%\begin{equation*}
%    \text{Recall}@k
%    = \frac{1}{|\vy|}\sum_{j \in \text{top}_k(\hat{\vy})} \vy_j
%\end{equation*}
where $\text{top}_k(\hat{\vy})$ denotes labels with the $k$ largest predicted scores.

\subsection{Implementation Details}

\paragraph{\bf Comparison Methods.}
Our \PEFA framework is applicable to any black-box \ERMs.
We applied \PEFA to various competitive ERMs, such as \SBERT~\citep{reimers2019sentence},
\DPR~\citep{karpukhin2020dense}, \MPNet~\citep{song2020mpnet}, \STfive~\citep{ni2022sentence} and \GTR~\citep{ni2022large}.
We also compare \PEFA with recent SoTA Sequence-to-Sequence (Seq2Seq) models, including Differentiable Search Index (\DSI)~\citep{tay2022transformer}, Search Engines with Autoregressive LMs (\SEAL)~\citep{bevilacqua2022autoregressive} and  Neural Corpus Indexer (\NCI~)\citep{wang2022neural}.

\paragraph{\bf Hyper-parameters}
PEFA have two hyper-parameters, the interpolation coefficient $\lambda$ in Eq.~\ref{eq:pefa-general} and the number of nearest neighbors $k$ in Eq.~\ref{eq:pefa-xl} used by \PEFAxl only.
We present ablation studies on hyper-parameters in Section~\ref{sec:exp-retrieval-ablations} where $\lambda=\{0.1, 0.3, 0.5, 0.7, 0.9\}$ and $k=\{16, 32, 64\}$.
For $\lambda=1.0$, \PEFA reduce back to its baseline \ERM.
We consider HNSW for \ANN search and set hyper-parameters according to existing work~\cite{nigam2019semantic,chang2021extreme,magnani2022semantic}.
At the index building stage, the maximum edge per node $M=32$ and the size of priority queue for graph construction $efC=500$.
At the online serving stage, the beam search width for graph search $efS=300$ .

% \paragraph{\bf Computing Environments}
% Experiments are conducted on an AWS x2idn.24xlarge CPU machine, with 96 vCPUs and 1,536 GiB memory.
% The only exception is to get the query and product embeddings, which required access to an AWS p4d.24xlarge GPU machine, with 96 vCPUs, 8 A100 Nvidia GPUs, and 1,152 GiB memory.

\subsection{Main Results}
\label{sec:exp-retrieval-main}

\paragraph{\bf \NQ}
In Table~\ref{tab:exp-NQ320k}, we applied \PEFAxs and \PEFAxl to fine-tuned \ERMs: \SBERT~\citep{reimers2019sentence}, \DPR~\citep{karpukhin2020dense}, \MPNet~\citep{song2020mpnet}, \STfive~\citep{ni2022sentence} and \GTR~\citep{ni2022large}.
These \ERMs were full parameter fine-tuned on the \NQ dataset.
The proposed \PEFA framework achieved significant improvement for a wide range of black-box \ERMs. 
The average gain of \PEFAxs over \ERMs are $+9.22\%$ and $+5.29\%$, for Recall@10 and Recall@100, respectively.
The average gain of \PEFAxl over \ERMs are $+11.33\%$ and $+5.20\%$ for Recall@10 and Recall@100, respectively.
For competitive \ERMs such as \MPNet~\cite{song2020large} and \GTR~\citep{ni2022large}, 
\textbf{\PEFA further outperform the previous SoTA Seq2Seq method, namely \NCI~\citep{wang2022neural}.}
The Recall@10 and Recall@100 of \MPNet+\PEFAxl are $88.72\%$ and $94.13\%$,
which is considerably better than the previous SoTA \NCI.
On the other hand, the original \MPNet without \PEFA can not outperform the previous SOTA  method, \NCI.
%especially for the Recall@10 metric.

\begin{table}[!ht]
\centering
\begin{tabular}{l|rr}
\toprule
Methods & Recall@10 & Recall@100 \\
\midrule
\BM                                                 & 32.48 & 50.54 \\
\DSI(base)~\citep{tay2022transformer}               & 56.60 & - \\
\NCI(base)~\citep{wang2022neural}                   & 85.20 & 92.42 \\
\SEAL(large)~\citep{bevilacqua2022autoregressive}   & 81.24 & 90.93 \\
\midrule

\SBERT~\citep{reimers2019sentence}           & 67.08 & 81.40 \\
\quad$+$\PEFAxs (ours)                       & 80.52 & 92.22 \\ \vspace{.1em}
\quad$+$\PEFAxl (ours)                       & 85.26 & 92.53 \\
\DPR~\citep{karpukhin2020dense}              & 70.68 & 85.19 \\
\quad$+$\PEFAxs (ours)                       & 83.45 & 92.22 \\ \vspace{.1em}
\quad$+$\PEFAxl (ours)                       & 84.65 & 92.07 \\
\MPNet~\citep{song2020mpnet}                 & 80.82 & 92.39 \\
\quad$+$\PEFAxs (ours)                       & 86.67 & \underline{94.53} \\ \vspace{.1em}
\quad$+$\PEFAxl (ours)                       & \textbf{88.72} & \textbf{95.13} \\
\STfive~\citep{ni2022sentence}               & 73.63 & 88.16 \\
\quad$+$\PEFAxs (ours)                       & 82.52 & 92.18 \\ \vspace{.1em}
\quad$+$\PEFAxl (ours)                       & 83.69 & 92.55 \\
\GTR~\citep{ni2022large}                     & 79.74 & 90.91 \\
\quad$+$\PEFAxs (ours)                       & 84.90 & 93.28 \\ \vspace{.1em}
\quad$+$\PEFAxl (ours)                       & \underline{88.71} & 94.36 \\
\midrule
Avg. Gain of \PEFAxs over \ERM               &  +9.22 & +5.28 \\
Avg. Gain of \PEFAxl over \ERM               & +11.82 & +5.72 \\
\bottomrule
\end{tabular}%
\caption{
    Our \PEFA framework on \NQ dataset.
    Both Seq2Seq models and \ERMs are full-parameter fine-tuned.
    The results of \BM, \DSI~\citep{tay2022transformer}, \NCI~\citep{wang2022neural} and \SEAL~\citep{bevilacqua2022autoregressive} are taken from ~\citep{wang2022neural}.
    1st/2nd place numbers are boldface/\underline{underscore}, respectively.
    %Best numbers are boldface and 2nd place numbers are underscore.
}
\label{tab:exp-NQ320k}
\vspace{-2.00em}
\end{table}

\paragraph{\bf \TriviaQA}
In Table~\ref{tab:exp-TriviaQA}, we applied \PEFAxs and \PEFAxl to pre-trained \ERMs.
Note that these \ERMs were not fine-tuned with any relevant query-document pairs from \TriviaQA.
The setup examines the robustness and generalization of our \PEFA framework.
We observe \PEFAxs and \PEFAxl achieve larger average gain of Recall over the unsupervised \ERMs,
when comparing Table~\ref{tab:exp-TriviaQA} to Table~\ref{tab:exp-NQ320k}.
When the underlying \ERM are pre-traiend only (not fine-tuned to the downstream task),
\PEFAxs seems to perform slightly better than \PEFAxl in Recall@20,
where the former has an average gain of $18.67\%$ while the latter
has an average gain of $17.07\%$.

\begin{table}[!ht]
\centering
\begin{tabular}{l|rr}
\toprule
Methods & Recall@20 & Recall@100 \\
\midrule
\BM                                                 & 69.45 & 80.24 \\
\NCI(base)~\citep{wang2022neural}                   & \textbf{94.45} & \textbf{96.94} \\
\SEAL(large)~\citep{bevilacqua2022autoregressive}   & 81.24 & 90.93 \\
\midrule

\SBERT~\citep{reimers2019sentence}           & 51.94 & 68.50 \\
\quad$+$\PEFAxs (ours)                       & \underline{86.28} & \underline{93.33} \\ \vspace{.1em}
\quad$+$\PEFAxl (ours)                       & 83.76 & 91.83 \\
\DPR~\citep{karpukhin2020dense}              & 60.69 & 73.80 \\
\quad$+$\PEFAxs (ours)                       & 82.97 & 91.06 \\ \vspace{.1em}
\quad$+$\PEFAxl (ours)                       & 78.76 & 89.62 \\
\MPNet~\citep{song2020mpnet}                 & 77.03 & 87.34 \\
\quad$+$\PEFAxs (ours)                       & 86.05 & 92.97 \\ \vspace{.1em}
\quad$+$\PEFAxl (ours)                       & 86.13 & 92.42 \\
\STfive~\citep{ni2022sentence}               & 62.74 & 77.21 \\
\quad$+$\PEFAxs (ours)                       & 78.39 & 88.57 \\ \vspace{.1em}
\quad$+$\PEFAxl (ours)                       & 75.13 & 87.24 \\
\GTR~\citep{ni2022large}                     & 71.75 & 82.05 \\
\quad$+$\PEFAxs (ours)                       & 83.81 & 91.02 \\ \vspace{.1em}
\quad$+$\PEFAxl (ours)                       & 85.30 & 92.38 \\
\midrule
Avg. Gain of \PEFAxs over \ERM               & +18.67 & +13.61 \\
Avg. Gain of \PEFAxl over \ERM               & +17.07 & +12.80 \\
\bottomrule
\end{tabular}%
\caption{
    Our \PEFA framework on \TriviaQA dataset.
    Seq2Seq models are full-parameter fine-tuned while \ERMs are unsupervisedly pre-trained.
    \ERMs+\PEFA did not fine-tune or update any parameter of the underlying \ERMs.
    1st/2nd place numbers are boldface/\underline{underscore}, respectively.
    %Best numbers are boldface and 2nd place numbers are underscore.
}
\vspace{-2em}
\label{tab:exp-TriviaQA}
\end{table}

\subsection{Ablation Studies}
\label{sec:exp-retrieval-ablations}

In Table~\ref{tab:exp-ablations},
we present ablation studies of two hyper-parameters of our \PEFA framework on the \NQ dataset.
$\lambda$ is the interpolation coefficient
that balances $f_{\text{\ERM}}$ and $f_{\text{\kNN}}$ in Equation~\ref{eq:pefa-general}.
When $0.0<\lambda<1.0$,
the Recall@100 of both \PEFAxs and \PEFAxl are consistently higher the \ERM alone ($\lambda=1.0$).
For \PEFAxs and \PEFAxl, $\lambda=0.5$ and $\lambda=0.1$ mostly yield the largest gain in average, respectively.
Crucially, the linear interpolation of \PEFAxs can be pre-computed offline at the HNSW index building stage (see Figure~\ref{fig:pefa-xs}) hence did not increase any inference latency overhead compared to the \ERMs.
For \PEFAxl, besides the hyper-parameter $\lambda$, it has another hyper-parameter $k$,
controlling the number of nearest neighbors in the \kNN model $f_{\text{\kNN}}$. 
We observed that $k=32$ generally saturate the performance.

% Ablation Table
\begin{table}[!ht]
\centering
\resizebox{0.975\linewidth}{!}{%
\begin{tabular}{c|l|rrrrr}
\toprule
\multirow{2}{*}{\ERM} 
    & \multicolumn{1}{|c|}{\multirow{2}{*}{\PEFA}}
    & \multicolumn{5}{c}{Recall@100 of various $\lambda$} \\
& & \multicolumn{1}{c}{0.1} & \multicolumn{1}{c}{0.3} & \multicolumn{1}{c}{0.5}  & \multicolumn{1}{c}{0.7} & \multicolumn{1}{c}{0.9} \\
\midrule
\multirow{4}{*}{\DPR}       & \PEFAxs          & 91.48 & 92.22 & 91.71 & 89.87 & 87.08 \\
                            & \PEFAxl (k=16)   & 91.98 & 90.66 & 89.72 & 88.54 & 87.62 \\
                            & \PEFAxl (k=32)   & 92.07 & 90.50 & 89.20 & 88.62 & 87.46 \\
                            & \PEFAxl (k=64)   & 91.93 & 89.89 & 88.95 & 88.39 & 87.04 \\ \midrule
\multirow{4}{*}{\STfive}    & \PEFAxs          & 91.23 & 92.16 & 92.20 & 91.61 & 89.72 \\
                            & \PEFAxl (k=16)   & 92.53 & 91.25 & 90.82 & 90.69 & 90.24 \\
                            & \PEFAxl (k=32)   & 92.34 & 91.20 & 90.96 & 90.77 & 90.11 \\
                            & \PEFAxl (k=64)   & 92.22 & 91.26 & 91.03 & 90.70 & 89.90 \\ \midrule
\multirow{5}{*}{\GTR}       & \PEFAxs          & 92.11 & 93.07 & 93.31 & 92.85 & 91.74 \\
                            & \PEFAxl (k=16)   & 94.36 & 93.32 & 92.81 & 92.53 & 91.93 \\
                            & \PEFAxl (k=32)   & 94.32 & 93.23 & 92.82 & 92.44 & 91.79 \\
                            & \PEFAxl (k=64)   & 93.93 & 93.14 & 92.76 & 92.29 & 91.62 \\ \bottomrule
\end{tabular}%
}
\caption{
    Ablation study of our \PEFA framework on \NQ dataset.
    \PEFAxs has only one hyper-parameter, namely the interpolation coefficient $\lambda$.
    \PEFAxl has two hyper-parameters: $\lambda$ and $k$ (number of nearest neighbors).
    For $\lambda=1.0$, \PEFAxs and \PEFAxl reduce to the same underlying \ERM in Table~\ref{tab:exp-NQ320k}.
}
\vspace{-1em}
\label{tab:exp-ablations}
\end{table}

%% file: contents/s5-exp-matching.tex
\section{Experiments on Product Search}
\label{sec:exp-prod}

For large-scale product search system, full-parameter fine-tuning may take thousands of GPU hours.
In this section, we conducted experiments on such larger-scale datasets and demonstrated that our \PEFA framework is an effective and fast technique that offers sizable improvements to not only a variety of pre-trained \ERMs but also the full-parameter fine-tuned \ERMs.

\subsection{Datasets \& Evaluation Protocols}
\label{sec:exp-prod-setup}

\paragraph{\bf Datasets}
We follow similar procedure ~\citep{nigam2019semantic,chang2021extreme,lu2021graph,hui2022augmenting}, to collect datasets from a large e-commerce product search engine.
Based on the size of catalog $n=|\gP|$, we construct three subsets as follows.
\begin{itemize}
    \item \DATAau: consists of roughly 30 millions of relevant query-product pairs, which covers around 10 millions of queries and 5 millions of products.
    \item \DATAmx: consists of roughly 150 millions of relevant query-product pairs, which covers around 40 millions of queries and 15 millions of products.
    \item \DATAca: consists of roughly 500 millions of relevant query-product pairs, which covers around 100 millions of queries and 30 millions of products.
\end{itemize}
For all proprietary ProdSearch datasets, the data statistics do not reflect the real traffic of the e-commerce system due to privacy concerns.
All relevant query-product pairs are random samples from anonymous aggregated search log.
We further split those pairs into the training set and the test set by time horizon,
where we use first twelve months of search logs as the training set and the last one month of search logs as the evaluation test set.

\begin{table*}[!ht]
\centering
\begin{tabular}{l|rr|rr|rr}
\toprule
\multirow{2}{*}{Methods}        & \multicolumn{2}{c|}{\DATAau} & \multicolumn{2}{c|}{\DATAmx} & \multicolumn{2}{c}{\DATAca} \\
                                & Recall@100 & Recall@1000 & Recall@100 & Recall@1000 & Recall@100 & Recall@1000 \\ \midrule
\MPNet~\citep{song2020mpnet}    &  0.00 &  0.00 &  0.00 &  0.00  &  0.00 &  0.00 \\
\quad$+$\PEFAxs (ours)           & 11.23 & 13.14 &  5.05 & 11.79  &  9.67 & 17.47 \\ \vspace{.25em}
\quad$+$\PEFAxl (ours)           & 22.83 & 12.31 & 23.48 & 21.56  & 27.22 & 18.96 \\
\STfive~\citep{ni2022sentence}  &  0.44 &  3.42 &  1.32 &  3.44  &  1.89 &  5.17 \\
\quad$+$\PEFAxs (ours)           & 13.63 & 17.13 & 10.39 & 16.34  & 13.18 & 21.28 \\ \vspace{.25em}
\quad$+$\PEFAxl (ours)           & 23.09 & 13.43 & 23.91 & 23.72  & 30.10 & 21.25 \\
\GTR~\citep{ni2022large}        &  7.85 &  9.23 &  6.75 & 10.33  &  8.35 &  9.83 \\
\quad$+$\PEFAxs (ours)           & 17.32 & 19.55 & 16.83 & 25.00  & 18.49 & 24.38 \\ \vspace{.25em}
\quad$+$\PEFAxl (ours)           & \underline{27.79} & 19.23 & 27.87 & 28.75  & 31.71 & 24.28 \\
\Efive~\citep{wang2022text}     &  9.93 &  9.75 &  9.98 & 12.98  & 12.01 & 12.61 \\
\quad$+$\PEFAxs (ours)           & 19.23 & 19.18 & 17.21 & 27.78  & 20.11 & 26.08 \\ \vspace{.25em}
\quad$+$\PEFAxl (ours)           & 26.83 & 17.75 & \underline{30.48} & 31.07  & \underline{31.91} & 25.49 \\ \midrule
\ProdERM~\citep{muhamed2023web} & 21.32 & 20.87 & 21.74 & 30.04  & 18.49 & 24.11 \\
\quad$+$\PEFAxs (ours)           & 23.42 & \underline{22.17} & 26.34 & \underline{34.84}  & 23.79 & \underline{29.61} \\ \vspace{.25em}
\quad$+$\PEFAxl (ours)           & \textbf{29.32} & \textbf{22.87} & \textbf{36.54} & \textbf{37.24}  & \textbf{32.99} & \textbf{30.01} \\ \midrule
Avg. Gain of \PEFAxs             & 16.97 & 18.23 & 15.16 & 23.15  & 17.05 & 23.76 \\
Avg. Gain of \PEFAxl             & 25.97 & 17.12 & 28.46 & 28.47  & 30.79 & 24.00 \\
\bottomrule
\end{tabular}%
\caption{
    Applying \PEFA to pre-trained \ERMs (\MPNet~\citep{song2020mpnet}, \STfive~\citep{ni2022sentence}, \GTR~\citep{ni2022large} and \Efive~\citep{wang2022text}) and the fine-tuned \ERMs (\ProdERM~\citep{muhamed2023web}) on three proprietary product search datasets: \DATAau, \DATAmx and \DATAca.
    To avoid disclosing the exact performance of production systems for privacy concerns,
    all reported numbers are absolute gain of Recall metrics compared to the baseline method \MPNet.
    1st/2nd place numbers are boldface/\underline{underscore}, respectively.
}
\vspace{-1em}
\label{tab:exp-prod-recall}
\end{table*}

\paragraph{\bf Evaluation Protocol}
%For the test set, we further sample up to 50,000 queries that have at least 1 purchased products.
To eliminate evaluation bias toward our \PEFA framework, all test queries are \textit{unseen} in the training set.
To avoid disclosing the exact performance of production systems, 
we report absolute gain of Recall@k metrics between the proposed \PEFA framework and the baseline \ERMs. 

We also report the ANN index size (GiB) and the index building time (hours) in offline indexing stage.
For online inference, following the ANN benchmark protocol~\citep{aumuller2020ann}, 
we consider the single thread setup and report the inference latency (milliseconds/query).

\subsection{Main Results}
\label{sec:exp-prod-main}

In Table~\ref{tab:exp-prod-recall}, we applied \PEFA to pre-trained \ERMs
(e.g., \MPNet~\citep{song2020mpnet}, \STfive~\citep{ni2022sentence}, \GTR~\citep{ni2022large} and \Efive~\citep{wang2022text})
and the fine-tuned \ERMs (\ProdERM~\citep{muhamed2023web}).
For privacy of the proprietary product search datasets,
we only report the absolute gain of Recall metrics compared to the \MPNet baseline.

Without \PEFA, pre-trained \ERMs have much lower Recall metrics compared to \ProdERM,
as the latter is carefully pre-trained and fine-tuned.
Adding \PEFAxs and \PEFAxl to those pre-trained \ERMs significantly lift the Recall
to comparable, or even outperform, the fine-tuned \ProdERM.
Take the largest dataset \DATAca as an example.
Adding \PEFAxl to \STfive, \GTR and \Efive have a Recall@100 gain of $30.10\%$, $31.71\%$ and $31.91\%$, respectively.
These recall@100 gain is already outperform the Recall@100 of fine-tuned \ProdERM. 
On the other hand, \PEFAxs on pre-trained \ERMs offer smaller Recall gain compared to \PEFAxl.
Only \Efive$+$\PEFAxs have a larger Recall@100 gain compared to the fine-tuned \ProdERM.

Similar to the finding of \NQ, we also see that \PEFA can further improve the performance of fine-tuned \ERMs.
For example, on the largest dataset \DATAca,
\PEFAxs and \PEFAxl further improve the Recall@100 of the fine-tuned \ProdERM by $5.3\%$ and $14.50\%$, respectively.

\subsection{Indexing and Inference}

In Table~\ref{tab:exp-prod-latency},
we discuss the trade-off between the performance and the deployment efficiency for the proposed \PEFA framework.
Note that \PEFA is a parameter-free method without updating model parameters of \ERMs, which can be easily implemented in the offline HNSW index building stage.
For the largest dataset \DATAca, 
the run-time of building HNSW indices for \PEFAxs and \PEFAxl  are $1.0$ and $4.7$ hours, respectively.
This is much faster than hundred of GPU hours when fine-tuning the \ProdERM on the billion-scale dataset.

\begin{table}[!ht]
\centering
\resizebox{\columnwidth}{!}{%
\begin{tabular}{c|l|rr|r}
\toprule
\multirow{2}{*}{Datasets} & \multirow{2}{*}{Methods} & \multicolumn{2}{c|}{Indexing} & Serving \\
                          &                          &  disk-size & run-time & Latency \\ \midrule
\multirow{3}{*}{\DATAau}& \ProdERM       &  13.1 & 0.3 & 0.82 \\
                        & \quad$+$\PEFAxs &  13.1 & 0.2 & 0.67 \\ \vspace{.25em}
                        & \quad$+$\PEFAxl &  32.2 & 0.7 & 2.15 \\ \midrule
\multirow{3}{*}{\DATAmx}& \ProdERM       &  28.6 & 0.6 & 0.91 \\
                        & \quad$+$\PEFAxs &  28.6 & 0.5 & 0.94 \\ \vspace{.25em}
                        & \quad$+$\PEFAxl & 100.7 & 1.9 & 1.94 \\ \midrule
\multirow{3}{*}{\DATAca}& \ProdERM       &  51.9 & 0.9 & 0.77 \\
                        & \quad$+$\PEFAxs &  51.9 & 1.0 & 0.71 \\ \vspace{.25em}
                        & \quad$+$\PEFAxl & 287.7 & 4.7 & 1.99 \\                        
\bottomrule
\end{tabular}%
}
\caption{
    For practical deployment consideration,
    we report the HNSW index size on disk (GiB) and the run-time (hours) of \PEFA during offline index building stage.
    We also report the inference latency (millisecond/query) of the HNSW index for online serving.
}
\label{tab:exp-prod-latency}
\end{table}

Despite larger gain in recall metrics,
\PEFAxl comes at the cost of larger HNSW index, longer index building time, and larger inference latency.
Specifically, the HNSW index size of \PEFAxl is $3.6$x larger than the HNSW index of \ERM,
as \PEFAxl requires two HNSW indices:
One HNSW index on the product embeddings $\rmP \in \sR^{n \times d}$,
while the other HNSW index on the training query embeddings $\rmQ \in \sR^{m \times d}$ for the \kNN modeling.
For product search datasets, the number of queries $m$ can be larger than the number of products $n$.
Due to similar reasons, the inference latency of \PEFAxl is $2.4$x larger than the latency of \ERM.

On the other hand, \PEFAxs not only achieves modest gains of recall metrics,
but also maintains the same deployment efficiency (e.g., HNSW index size and inference latency) as its baseline \ERM.
Recall that \PEFAxs maintains only one ANN index because the interpolation of $f_{\text{\ERM}}$ and $f_{\text{\kNN}}$ is independent to test-time query, which can be pre-computed offline in a single ANN index (see Equation~\ref{eq:pefa-xs} in Section~\ref{sec:pefa-xs}).
From the deployment perspective,
\PEFAxs may be a more practical choice as it introduces zero additional overhead to the production system at inference time.

\subsection{Effect of Supervised Data Size}
\label{sec:exp-prod-dsize}

\begin{figure}[!h]
    \centering
    \begin{subfigure}{.235\textwidth}
        \centering
        \includegraphics[width=.975\linewidth]{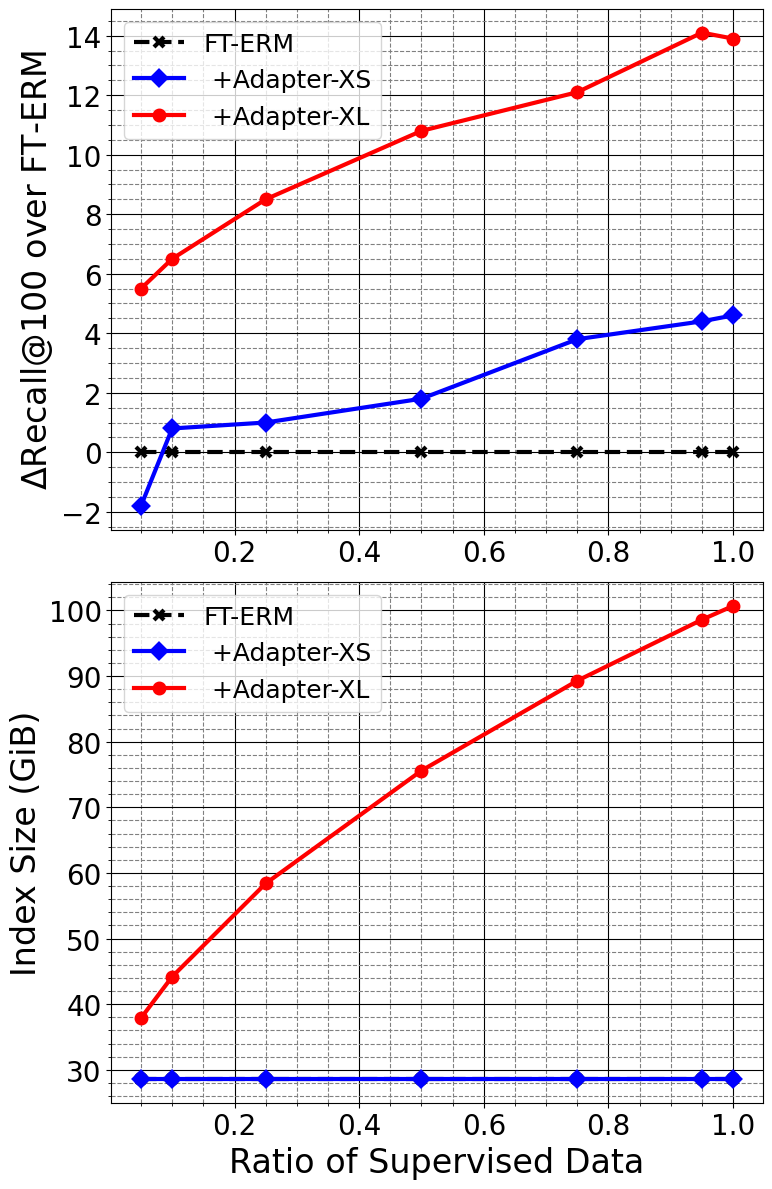}
        \caption{\DATAmx}
        \label{fig:exp-prod-dsize-mx}
    \end{subfigure}
    \begin{subfigure}{.235\textwidth}
        \centering
        \includegraphics[width=.975\linewidth]{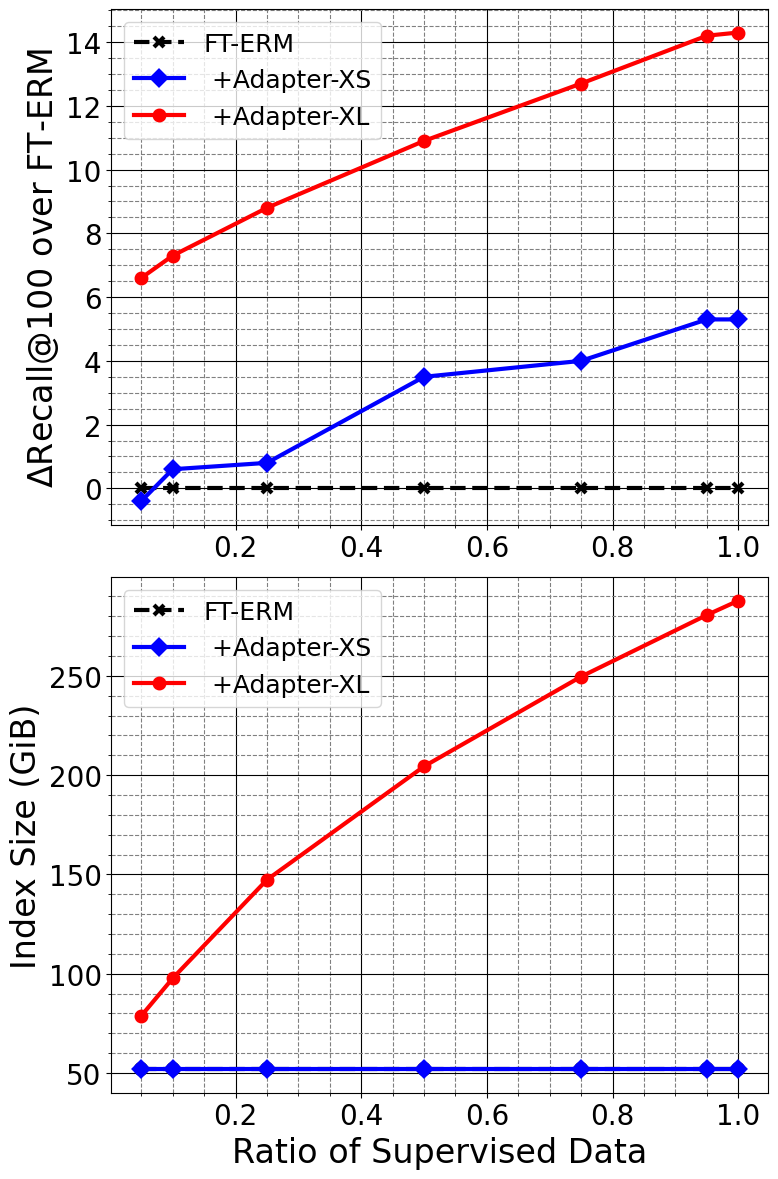}
        \caption{\DATAca}
        \label{fig:exp-prod-dsize-ca}
    \end{subfigure}
    \caption{
        The amount of supervised data versus predictive power and model size for the proposed \PEFA framework.
        The y-axis of the 1st row figures is the recall gain compared to the \ProdERM (black line).
        The y-axis of the 2nd row figures is the \PEFA model size (GiB).
        The x-axis of all figures is the ratio of supervised data being used.
    }
    \label{fig:exp-prod-dsize}
\end{figure}

The amount of supervised data (i.e., relevant query-product pairs) consumed by \PEFA 
plays a crucial role to the predictive power of \PEFAxs and \PEFAxl, run-time of HNSW index building, and the model size of resulting HNSW indices.
Hence, we present such analysis in Figure~\ref{fig:exp-prod-dsize}.
The amount of supervised data is controlled by the sampling ratio $\{0.05, 0.10, 0.25, 0.50, 0.75, 0.95\}$.
In particular, we uniformly sample query-product pairs
from the relevance matrix $\rmY \in \{0,1\}^{n \times m}$ in Equation~\ref{eq:knn-mat}.

With 10\% of the relevant query-product pairs sampled from $\rmY$,
\PEFAxs reaches the same level of Recall@100 compared to the fine-tuned \ProdERM.
With more supervised data, \PEFAxs outperforms \ProdERM eventually.
What's more, the model size of \PEFAxs do not increase as it consumes more supervised data.

For \PEFAxl, interestingly, it can achieve significant improvements in Recall@100 even with just $5\%$ of the supervised data.
At $5\%$ of the supervised data usage,
the resulting HNSW index is around $1.4$x times larger than the HNSW index of \ProdERM and \PEFAxs.
Also, the inference latency of \PEFAxl seems to be consistently $2$x larger than the latency of \ProdERM and \PEFAxs across all datasets.
Again, it is up to the practitioners to decide the trade-off between the additional performance gain brought by \PEFAxl and the cost of larger index size and inference latency.

%% file: contents/s6-relatedwork.tex
\section{Related Work}
\label{sec:relatedwork}

\subsection{Dense Text Retrieval}

%Shifting from traditions using sparse inverted index like BM25~\cite{manning2008introduction}, in the deep learning era, dense text retrieval~\cite{karpukhin2020dense} and embedding-based retrieval models (ERMs) have become one of the most popular approaches for search in industry because of its ease for production.
%With the advent of deep learning, dense text retrieval~\cite{karpukhin2020dense} and embedding-based retrieval models (ERMs) have become increasingly popular in industry for search, supplanting traditional approaches such as BM25~\cite{manning2008introduction} which use sparse inverted indexes. This is due in part to their ease of deployment in production settings.
%\ERMs learn the query and label representations within the same low-dimensional embedding space, so the relevance to a given query can be estimated by the dot-product score for each label.
%The inference can then be easily achieved by approximate nearest neighbor~(ANN) search algorithms like HNSW~\cite{malkov2018efficient} and Product Quantizations~\citep{johnson2019billion,guo2020accelerating}.
%To learn embedding vectors,
DSSM~\cite{huang2013learning} and C-DSSM~\cite{shen2014learning} utilize multi-layer perceptron and convolutional neural networks while
DPR~\cite{karpukhin2020dense} deploys pre-trained neural language models~(NLMs) like BERT~\cite{devlin2018bert}.
Some studies attempt to improve ERMs by pre-training and adjusting results.
Condenser~\cite{gao2021condenser} pre-trains NLMs with the idea of Funnel-Transformer~\cite{dai2020funnel} while Co-Condenser~\cite{gao2022unsupervised} re-ranks its retrieval results with an attentive cross-encoder.
DPTDR~\cite{tang2022dptdr} applies prompt-tuning~\cite{liu2023pre} to further improve the quality dual encoders for ERMs.
However, conventional ERMs could suffer from dealing with tail queries and labels, especially when we have an enormous industry-scale index~\cite{reimers2021curse}.
Even though some lines of research attempt to address this issue by computing label-centric similarity~\cite{ren2021pair} and multi-view representations~\cite{zhang2022multi}, they are infeasible for industrial production due to the requirement of extensive pre-training and additional cross-attentive computations between queries and labels.

Different from existing approaches, our \PEFA has no need of pre-training another embedding model, so the enhancement can be achieved within an acceptable short period of only ANN search indexing.
Moreover, the kNN model of training queries can further benefit the representation capability of \PEFA embedding.

\subsection{Inference with Training Instances}

Similar to our proposed \PEFA framework, some studies also leverage training instances in inference for better performance in various research fields.
$k$ nearest neighbors language modeling (\kNN-LM)
~\citep{khandelwal2020Generalization,yogatama2021adaptive,he2021efficient} build a kNN model within the small vocabulary space
while PIFA embeddings~\citep{jain2019slice,chang2021extreme,zhang2021fast,yu2022pecos} aggregate sparse representations of training instances for the label space.
The derived coresets of training embeddings can be indexed with ANN search to shrink candidate labels and accelerate inference for recommender systems~\cite{jiang2020clustering} and neural language models~\cite{chen2019learning}.
However, none of the above methods addresses the challenge of large-scale retrieval in an industry scale.

\subsection{Parameter-Efficient Tuning of \ERMs}
To avoid the expensive full-parameter fine-tuning of \ERMs for various downstream tasks,
there are some preliminary studies on parameter-efficient fine-tuning of \ERMs~\citep{jung2022semi,ma2022scattered,pal2023parameter}.
Nevertheless, as pointed out by~\citep{ma2022scattered},
naively apply existing parameter-efficient fine-tuning methods in the NLP literature,
such as Adapter~\citep{houlsby2019parameter}, prefix-tuning~\cite{li2021prefix} and LoRA~\citep{hu2022lora},
often results in limited success for \ERM in the retrieval applications.
Furthermore, parameter-efficient fine-tuning approaches still require access to the models' gradient, which may not be available for the recent powerful large language models (LLMs) such as GPT-3~\citep{brown2020language}.
Our proposed \PEFA framework is complementary to any pre-trained and fine-tuned \ERMs,
namely including \ERMs derived from parameter-efficient fine-tuning.
Notice that our \PEFA did not require any gradient information of the underlying \ERMs, which can have a broader impact to black-box \ERMs where the encoders are initialized from LLMs.

%% file: contents/s7-conclusion.tex
\section{Conclusions}
\label{sec:conclusions}

In this paper, we propose \PEFA, \textit{parameter-free} adapters for fast tuning of black-box \ERMs.
\PEFA offers flexible choices (i.e., \PEFAxs and \PEFAxl) for practitioners to improve their pre-trained or fine-tuned \ERMs efficiently, without any updates to model parameters of \ERMs.
\PEFAxl brings more significant gain of Recall@k at the cost of doubling the ANN index size and inference latency,
while \PEFAxs yields modest gain of Recall@k without any overhead compared to the existing \ERM inference pipeline.
For document retrieval, \PEFA not only improves the recall@100 of pre-trained \ERMs on \TriviaQA by an average of 13.2\%,
but also lifts the recall@100 of fine-tuned \ERMs on \NQ by an average of 5.5\%.
For \NQ dataset, applying \PEFA to \MPNet~\citep{song2020mpnet} and \GTR~\citep{ni2022large} reaches new SoTA results, where the Recall@10 of $88.72\%$ outperforms $85.20\%$ of previous SoTA Seq2Seq-based \NCI~\citep{wang2022neural}.
For product search consisting of billion-scale of data,
\PEFA improves the Recall@100 of the fine-tuned \ERMs
by an average of 5.3\% and 14.5\%, for \PEFAxs and \PEFAxl, respectively.